\documentclass[a4paper,12pt]{article}
\usepackage{amsmath,amsthm,amssymb,bbm}
\usepackage{xcolor}

\usepackage[english]{babel}
\usepackage{graphicx}

\DeclareMathOperator{\e}{e}

\title{\bf Local generation of entanglement with Redfield dynamics}

\author{F. Benatti$^{a,b}$, D. Chru\'sci\'nski$^{c}$, R. Floreanini$^{b}$ \\
\\
\small ${}^a$Dipartimento di Fisica, Universit\`a di Trieste, 34151 Trieste, Italy\\
\small ${}^b$Istituto Nazionale di Fisica Nucleare, Sezione di Trieste, 34151 Trieste, Italy\\
\small ${}^c$Institute of Physics, Faculty of Physics, Astronomy and Informatics,\\
\small Nicolaus Copernicus University, Grudziadzka 5/7, 87-100 Toru\'n, Poland
}

\date{\null}

\begin{document}

\maketitle

\begin{abstract}
\noindent
In phenomenological applications, time evolutions of Bloch-Redfield type are widely adopted for modelling
open system dynamics, despite their non-positive preserving character: this physical inconsistency,
that in general shows up at small times, is usually cured by suitably restricting the space of allowed initial states.
Nevertheless, additional problems may arise in relation to entanglement:
specifically, we show that Redfield dynamics can generate entanglement through a purely local action,
and this unphysical effect can persist for finite times.
\end{abstract}

\vskip 1cm

\section{Introduction}

Quantum systems immersed in large environments, typically heat baths, represent a paradigmatic
framework for modelling quantum non-unitary dissipative dynamics in phenomenological
applications. The time-evolution of such quantum systems can be obtained 
from the global system+environment dynamics by tracing over the environment (infinite) 
degrees of freedom and generally encodes decoherence and dissipation, including
possible memory effects \cite{Alicki}-\cite{Breuer}.

The derivation of such reduced time-evolution for the system alone from the microscopic
system-environment interaction Hamiltonian is however notoriously tricky, leading to a pletora
of different master equations, often generating physically inconsistent dynamics.
In particular, even in the Markovian limit, obtained when the system-environment coupling
is sufficiently weak and the environment correlations decay times are small with respect
to the characteristic time evolution of the system, the positivity of the system
density matrix might not be in general preserved \cite{Bloch}-\cite{Slichter}, with the remarkable exception
of {\it quantum dynamical semigroups}, whose derivation is however based on a rigorous 
mathematical treatment \cite{Alicki}-\cite{Rivas}.

Although acknowledged in the literature, these inconsistencies have been either dismissed 
as irrelevant for practical purposes or cured by adopting {\it ad hoc} prescriptions \cite{Gnutzmann}-\cite{Becker}.
Indeed, time-evolutions of the so-called Bloch-Redfield type \cite{Bloch}-\cite{Slichter}
are constantly used in applications, despite being non-positive \cite{Pollard}-\cite{Strunz}.%
\footnote{This may lead to claim the existence of physical 
results that are instead the consequence of the non-positivity of the
dynamics; see the discussion in \cite{Cattaneo}, in relation to \cite{Guarnieri}.}
As justification for this attitude, two facts are usually
remarked: {\it i)} non-positivity is in general confined to small times, and {\it ii)}
asymptotic thermalization is always guaranteed for such dynamics, 
while this is not always true for quantum dynamical semigroups, due to the constraints
imposed by complete positivity, which however guarantees physical consistency in all situations.

In the following, we shall consider Markovian reduced dynamics of Redfield type in the case of a
two-level system and point out further difficulties
of such time-evolutions related to the presence of entanglement. More specifically,
we shall study how the single qubit Redfield dynamics $\gamma_t$ behaves when augmented to
a factorized evolution $\Gamma_t= {\rm id} \otimes \gamma_t$, describing the dynamics
of the same qubit statistically coupled to a second ancillary qubit, which however remains
completely inert in time, being subjected to the identity operation ``id''.
We find that the purely local, factorized evolution $\Gamma_t$ can increase the entanglement of the two-qubit system,
clearly an unphysical result.

Pleminary studies on these topics have been previously reported in \cite{Benatti-2}-\cite{Benatti-4}, 
but limiting the considerations
to small times and very specific models. Instead, general Redfield evolutions are here examined,
explicitly showing that their inconsistencies in connection with entanglement generation
are not a small-time effect, rather they remarkably persist for finite times. 
As a result, the use of Redfield type time-evolutions for modelling open system dynamics
should be taken with great care.

\section{Redfield dynamics}

As previously remarked, we shall study the dynamics of a two-level system (qubit)
immersed in an environment, modelled as a large heat bath in equilibrium at the inverse temperature~$\beta$.
Being (infinitely) large, the reservoir can be considered unaffected by the presence of the qubit
and therefore in the reference Gibbs state:
\begin{equation}
\rho_B= \frac{ \e^{-\beta H_B} }{ {\rm Tr}_B [ \e^{-\beta H_B}]  }\ ,
\label{1.1}
\end{equation}
where $H_B$ is the Hamiltonian describing the bath dynamics.

On the other hand, in absence of the bath,
the qubit dynamics is driven by a $2\times 2$
Hamiltonian matrix $H_S$, that  can be taken
to assume the most general form:
\begin{equation}
H_S=\omega\, \vec n\cdot\vec\sigma\ ,
\label{1.2}
\end{equation}
where $\sigma_i$, $i=1,2,3$ are the Pauli matrices, $n_i$, $i=1,2,3$
are the components of a three-dimensional unit vector, 
while $2\omega$ represents the gap between the two qubit energy levels. 

Within the standard open system paradigm \cite{Alicki}-\cite{Rivas}, the interaction of the qubit 
with the bath is assumed to be weak and describable by
a Hamiltonian $H'$ that is linear in both qubit and environment variables:
\begin{equation}
H'=\sum_{i=1}^3\sigma_i\otimes B_i\ ,
\label{1.3}
\end{equation}
where $B_i$ are suitable hermitian bath operators.

The total Hamiltonian $H$ describing the complete system, the two-level system
together with the heat bath, can thus be written as
\begin{equation}
H=H_S\otimes {\bf 1}_B+ {\bf 1}_S\otimes H_B+ \lambda\, H'\ ,
\label{1.4}
\end{equation}
with $\lambda$ a small coupling constant.
It generates the time-evolution of the  total density matrix $\rho_{\rm tot}$, 
via the Liouville--von Neumann equation
\begin{equation}
{\partial\rho_{\rm tot}(t)\over\partial t}=\, 
-i[H,\, \rho_{\rm tot}(t)]\ ,
\label{1.5} 
\end{equation}
starting at $t=\,0$ from the initial
configuration $\rho_{\rm tot}(0)=\rho(0)\otimes \rho_B$, in absence of initial system-environment correlations.

Because of the weak coupling assumption, the dynamics of the reduced density matrix 
$\rho(t)\equiv{\rm Tr}_B[\rho_{\rm tot}(t)]$ for the qubit
is usually obtained through standard second-order approximations in the coupling $\lambda$
and a {\it naive} Markovian limit. In this way, one finds that the qubit time-evolution is generated
by a master equation of Bloch-Redfield type \cite{Bloch}-\cite{Slichter}:
\begin{equation}
{\partial\rho(t)\over\partial t}=\, 
-i[H_S,\, \rho(t)] + \mathbb{D}[\rho(t)]\ ,
\label{1.6}
\end{equation}
with the disipative part explicitly given by \cite{Benatti-1}:
\begin{equation}
\mathbb{D}[\rho]=- \lambda^2\int_0^{\infty}{\rm d}s\, 
{\rm Tr}_B\Bigl(
\Bigl[H'(s)\,,\,\Bigl[H'\,,\,\rho\otimes\rho_B\Bigr]\Bigr]\Bigr)\ .
\label{1.7}
\end{equation}
where
\begin{equation}
H'(s)=\sum_{i=1}^3\sigma_i(s)\otimes B_i(s)\ ,
\label{1.8}
\end{equation}
with qubit and bath operators following their free evolution,
\begin{eqnarray}
\label{1.9a}
&& \sigma_i(s)= {\rm e}^{isH_S}\,\sigma_i\,{\rm e}^{-isH_S}=\sum_{j=1}^3 U_{ij}(s)\, \sigma_j\ ,\\
&& B_i(s)={\rm e}^{isH_B}\,B_i\,{\rm e}^{-isH_B} \ ,
\label{1.9b}
\end{eqnarray}
and
\begin{equation}
U_{ij}(s)=n_i\, n_j +(\delta_{ij}-n_i\, n_j)\, \cos(2\omega s)
-\varepsilon_{ijk} n_k\, \sin(2\omega s)\ .
\label{1.10}
\end{equation}
By introducing the bath two-point correlation functions,
\begin{equation}
G_{ij}(s)={\rm Tr}_B\Bigl[\rho_B\, B_i(s)B_j\Bigr]=\Big[ G_{ji}(-s)\Big]^\dagger\ ,
\label{1.11}
\end{equation}
one can rewrite $\mathbb{D}$ in the more explicit form:
%
%
\begin{equation}
\mathbb{D}[\rho]=\,-\,\lambda^2
\sum_{i,j}\int_0^{\infty}{\rm d}s\Biggl\{
G_{ij}(s)\,
\Bigl[\sigma_i(s)\,,\,\sigma_j\,\rho\Bigr]\\
+G_{ji}(-s)\,
\Bigl[\rho\,\sigma_j\ ,\,\sigma_i(s)\Bigr]\Biggr\}\ .
\label{1.12}
\end{equation}
Using (\ref{1.9a}) and (\ref{1.10}), this expression can be further brought to Kossakowski-Lindblad form:
\begin{equation}
\mathbb{D}[\rho]= -i\Big[ H_{LS}, \rho\Big]
+\sum_{i,j=1}^{3}{\cal C}_{ij}\Bigl(\sigma_j\rho\,\sigma_i-
{1\over 2}\Bigl\{\sigma_i\sigma_j,\, \rho\Bigr\}\Bigr)\ ,
\label{1.13}
\end{equation}
with
\begin{equation}
H_{LS}=\frac{\lambda^2}{2}\sum_{i,j=1}^{3}\varepsilon_{ijk}\big(C_{ij}-C_{ji}^*\big)\, \sigma_k\ ,\qquad
{\cal C}_{ij}=\lambda^2\, \big(C_{ij}+C_{ji}^*\big)\ ,
\label{1.14}
\end{equation}
and
\begin{equation}
C_{ij}=\sum_{k=1}^{3} \int_0^{\infty}{\rm d}s \ U_{ki}(s)\, G_{kj}(s)\ .
\label{1.15}
\end{equation}
The first contribution in (\ref{1.13}) is of Hamiltonian form, the so-called {\it Lamb-shift},
that ``renormalizes'' the starting system Hamiltonian (\ref{1.2}); instead, the second contribution
is a purely dissipative one.

When written in this form, one realizes that the Redfield reduced time evolution $\gamma_t$ 
generated by (\ref{1.6}), (\ref{1.13}) would be a physically consistent dynamics, provided the (hermitian) 
Kossakowski matrix ${\cal C}_{ij}$ is non-negative; indeed, in this case, $\gamma_t$ would be
a semigroup of completely positive maps. Unfortunately, because of the form (\ref{1.15})
and the presence of the trigonometric functions in (\ref{1.10}), in general ${\cal C}_{ij}$ possesses
negative eigenvalues leading to finite dynamics $\gamma_t$ that often are not even positive.

In order to explicitly expose such inconsistencies, it is convenient to choose the unit vector 
$\vec n$ in (\ref{1.2}) to point in the third direction, and assume that the heat bath is such
that the correlation matrix (\ref{1.11}) is diagonal, with only the elements $G_{11}$, $G_{22}$, $G_{33}$ non-zero.%
\footnote{This choice is not much restrictive and it will be further discussed in the next Section.}
In this case, the Kossakowski matrix ${\cal C}_{ij}$ takes the form
\begin{equation}
{\cal C}=
\begin{pmatrix}
{\cal C}_{11} & {\cal C}_{12} & 0\\
{\cal C}_{21} & {\cal C}_{22} & 0\\
0 & 0 & {\cal C}_{33}
\end{pmatrix} \ ,
\label{1.16}
\end{equation}
with
\begin{eqnarray}
\nonumber
&& {\cal C}_{11} = \frac{\lambda^2}{2} \int_{-\infty}^\infty\, {\rm d}s \e^{2i\omega s}\, \Big( G_{11}(s) + G_{11}(-s)\Big)\ ,\\
\nonumber
&& {\cal C}_{22} = \frac{\lambda^2}{2} \int_{-\infty}^\infty\, {\rm d}s \e^{2i\omega s}\, \Big( G_{22}(s) + G_{22}(-s)\Big)\ ,\\
\label{1.17}
&& {\cal C}_{12} = \lambda^2 \int_0^\infty\, {\rm d}s\, \sin(2\omega s)\, \Big( G_{22}(s) - G_{11}(-s)\Big) = \big[{\cal C}_{21}\big]^*\ ,\\
\nonumber
&& {\cal C}_{33} =\lambda^2 \int_{-\infty}^\infty\,{\rm d}s \,  G_{33}(s)\ .\\
\nonumber
\end{eqnarray}
Similarly, the Lamb-shift contribution in (\ref{1.14}) becomes proportional to $\sigma_3$,
\begin{equation}
H_{LS}=\frac{\lambda^2}{2}\, \delta\omega\, \sigma_3\ ,\qquad
\delta\omega=\int_0^\infty\, ds \sin(2\omega s)\, \Big( G_{11}(s) + G_{11}(-s) + G_{22}(s) + G_{22}(-s) \Big)\ ,
\label{1.18}
\end{equation}
so that, due to the presence of the heat bath, the initial qubit frequency $\omega$ gets a 
$\lambda^2$-dependent shift:
\begin{equation}
\omega \to \tilde\omega = \omega + \frac{\lambda^2}{2}\, \delta\omega\ .
\label{1.19}
\end{equation}

As we are dealing with a two-dimensional system, it proves convenient to adopt a
vector-like representation by decomposing the qubit density matrix as
\begin{equation}
\rho=\frac{1}{2}\Big(\sigma_0+\vec{\rho}\cdot\vec{\sigma}\Big)=
\frac{1}{2}
\begin{pmatrix}
1+\rho_3&\rho_1-i\rho_2\\
\rho_1+i\rho_2&1-\rho_3\\
\end{pmatrix}
\ ,\qquad
{\rm Det}[\rho]=\frac{1}{4}\biggl(1-\sum_{j=1}^3\rho_j^2\biggr)\geq0\ ,
\label{1.20}
\end{equation}
where $\sigma_0$ is the two-dimensional unit matrix, while $\vec{\rho}$ 
is the three-dimensional coherence or Bloch vector, of unit length 
for pure states.
By representing $\rho$ as a $4$-vector 
$\vert\rho\rangle \equiv(1,\rho_1,\rho_2,\rho_3)$, the evolution equation (\ref{1.6}), (\ref{1.13}) can then be recast in a
Schr\"odinger-like form 
\begin{equation}
\partial_t\vert\rho(t)\rangle=\,-2\, \mathcal{L}\, 
\vert\rho(t)\rangle\ ,
\label{1.21}
\end{equation}
where the $4\times 4$ matrices $\mathcal{L}$ includes both the 
Hamiltonian and dissipative contributions:
\begin{equation}
\mathcal{L} =
\begin{pmatrix}
0 & 0 & 0 & 0\\
0 & a & b+\tilde\omega & 0\\
0 & b-\tilde\omega & \alpha & 0\\
w & 0 & 0 & \gamma\\
\end{pmatrix}\ ,
\label{1.22}
\end{equation}
through $\tilde\omega$ in (\ref{1.19}) and the real parameters:
\begin{equation}
\begin{array}{ll}
a= {\cal C}_{22} + {\cal C}_{33}\ , & \quad b= -{\cal R}e\big( {\cal C}_{12} \big)\ ,\\
\alpha= {\cal C}_{11} + {\cal C}_{33}\ , & \quad w=-2\,{\cal I}m\big( {\cal C}_{12} \big)\ ,\\
\gamma={\cal C}_{11} + {\cal C}_{22}\ .\\
\end{array}
\label{1.23}
\end{equation}
These parameters are not completely arbitrary. Since the environment state $\rho_B$ is thermal,
the correlations in (\ref{1.11}) obey the so called {\it Kubo-Martin-Schwinger (KMS) condition} \cite{Thirring,Bratteli}:
\begin{equation}
G_{ij}(t) = G_{ji}(-t-i\beta)\ ,
\label{1.24}
\end{equation}
that expresses the analiticity properties of thermal
correlation functions with respect to time; it can be easily derived
when $H_B$ has discrete spectrum, but survive the
thermodynamic limit and thus holds also for truly infinite environments.
From it, and the explicit expressions (\ref{1.17}), one then easily deduces the following relation:
\begin{equation}
\gamma-w = \e^{-2\beta\omega}\, (\gamma + w)\ ,
\label{1.25}
\end{equation}
and thus, assuming $\gamma$ nonvanishing,
\begin{equation}
\frac{w}{\gamma} = \frac{1-\e^{-2\beta\omega}}{1+\e^{-2\beta\omega}}\ .
\label{1.26}
\end{equation}
In addition, as expressible in terms of positive combination of Fourier transform of two-point bath
correlations, the parameters $a$, $\alpha$ and $\gamma$ turn out to be nonnegative ({\it e.g.} see \cite{Alicki}).

In this particular case, the solution of (\ref{1.21}) can be straightforwardy computed and the finite
time evoulution map $\gamma_t=\e^{-2t\mathcal{L}}$ explicitly obtained; starting from the initial
values, $\rho_1$, $\rho_2$ and $\rho_3$, the components
of the Bloch vector evolve in time according to:
\begin{eqnarray}
\nonumber
&& \rho_1(t) = M_{11}(t)\, \rho_1 + M_{12}(t)\, \rho_2\ ,\\
\label{1.27}
&& \rho_2(t) = M_{21}(t)\, \rho_1 + M_{22}(t)\, \rho_2\ ,\\
\nonumber
&& \rho_3(t)= \e^{-2\gamma\, t} \rho_3 +\Lambda(t)\ ,\\
\nonumber
\end{eqnarray}
where
\begin{eqnarray}
\nonumber
&& M_{11}(t) = \e^{-(a+\alpha) t}\ \bigg[\cos \big(\Omega t\big) - \frac{a-\alpha}{\Omega}\, \sin\big(\Omega t\big)\bigg]\ ,\\
\nonumber
&& M_{22}(t) = \e^{-(a+\alpha) t}\ \bigg[\cos \big(\Omega t\big) + \frac{a-\alpha}{\Omega}\, \sin\big(\Omega t\big)\bigg]\ ,\\
\label{1.28}
&& M_{12}(t) = -2 \e^{-(a+\alpha) t}\ \bigg[\frac{b+\tilde\omega}{\Omega}\, \sin\big(\Omega t\big)\bigg]\ ,\\
\nonumber
&& M_{21}(t) = -2 \e^{-(a+\alpha) t}\ \bigg[\frac{b-\tilde\omega}{\Omega}\, \sin\big(\Omega t\big)\bigg]\ ,\\
\nonumber
&& \Lambda(t)=-\frac{w}{\gamma}\Big( 1- \e^{-2\gamma\, t}\Big)\ ,\\
\nonumber
\end{eqnarray}
while
\begin{equation}
\Omega=\big(4\,\tilde\omega^2 - 4\, b^2 - (a-\alpha)^2\big)^{1/2}\ ,
\label{1.29}
\end{equation}
is a positive frequency, as $\lambda\ll\omega$.
From these explicit expressions, one immediately deduces that the Redfield dynamics asymptotically drives
the system to the unique equilibrium state $\rho_\infty$, with Bloch vector components
$\vec\rho_\infty=(0,0,-w/\gamma)$. Recalling (\ref{1.26}), one immediately finds
\begin{equation}
\rho_\infty=\frac{{\rm e}^{-\beta H_S}}{{\rm Tr}[{\rm e}^{-\beta H_S}]}\ ,
\label{1.30}
\end{equation}
so that the bath drives the qubit to an equilibrium Gibbs state
at the bath temperature. As mentioned before, this characteristics of the Redfield dynamics
makes it very appealing in applications.

On the other hand, the evolution (\ref{1.27}) does not in general preserve the positivity of the qubit 
density matrix $\rho(t)$ for all times. Indeed, let us consider the derivative of ${\rm Det}[\rho(t)]$
at the initial time $t=\,0$:
\begin{equation}
\frac{{\rm d}}{{\rm d}t}\hbox{Det}[\rho(t)]\Big|_{t=\,0}=2\Big[ a\rho_1^2 + \alpha \rho_2^2 +2b\rho_1\rho_2
+\rho_3\big(w +\gamma\rho_3\big)\Big]\ .
\label{1.31}
\end{equation}
For a pure initial state, such that $\rho_3=\,0$, $\rho_1^2 + \rho_2^2=1$ so that $\hbox{Det}[\rho]=\,0$
(recall (\ref{1.20})), this derivative must be positive, otherwise negative probabilities
would emerge as soon as the dynamics $\gamma_t$ starts. However, the quadratic form
$a\rho_1^2 + \alpha \rho_2^2 +2b\rho_1\rho_2$ in (\ref{1.31}), with coefficients as in (\ref{1.23}), 
need not be positive, so that indeed the dynamical map $\gamma_t$ turns out to be in general non-positive.

It is interesting to notice that the origin of the lack of positivity-preservation of
Redfiel dynamics lies in the way the
time-evolution generated by the free system Hamiltonian $H_S$ interferes with the various
approximations. Indeed, in absence of the free system evolution, $\omega=\,0$,
from (\ref{1.17}) one immediately obtains $w=b=0$. However, in this case the dynamical map $\gamma_t$ 
turns out to be completely positive and not just simply positive.

As a cure to the appearance of negative probabilties, it has been proposed
to restrict the space of initial conditions to those states $\rho$ that remain positive
under the action of the Redfield dynamics \cite{Gnutzmann}-\cite{Wielkie}. 
The general argument supporting this choice is that
negative probabilities appear only at the start of the evolution, {\it i.e.} at
short transient times, before the truly Markovian regime sets in, and therefore
in a span of time not really covered by the Redfield approximation.
However, as we shall see below, Redfield dynamics may be affected by additional,
more serious inconsistencies in presence of entanglement when dealing with 
multi-partite systems.

\section{Redfield dynamics and entanglement}

Let us now extend the treatment discussed in the previous section
to a bipartite system, composed by two independent qubits, the first one
just an inert ancilla, while the second one subjected to
the action of a heat bath and evolving with the previously considered Redfield dynamics $\gamma_t$.
The time evolution of the compound two-qubit system is then given by
$\Gamma_t={\rm id}\otimes\gamma_t$, where ``id'' represents the identity map.
As initial two-qubit state, we shall choose an entangled state and follow its evolution
under the dynamical map $\Gamma_t$. 

Although this setting might at first sight appear artificial, the situation where a qubit, subjected to a noisy
environement, is statistically correlated to another independent and dynamically inert ancilla 
is common in quantum information: it is the physical context where an entangled two-qubit state ({\it e.g.} a Bell state)
is formed in the laboratory and, while one qubit is kept inert there, the second one is sent to
another party via a noisy channel.

As the dynamics $\Gamma_t$ acts locally on the two parties, no interaction between the two qubits
is at work, a steady depletion of entanglement is expected. Instead, even starting from initial states
whose positivity is preserved by $\Gamma_t$ and whose reduced, one-qubit state remains positive
under the Redfield dynamics $\gamma_t$, we shall see that the dynamical
map $\Gamma_t$ is able to periodically increase the two-qubit entanglement, even at finite times,
clearly an unphysical behaviour.

In order to simplify the treatment, we shall limit our considerations to a special class of
two-qubit density matrices, those with non-vanishing entries only along the two diagonals
(in the two-qubit computational basis):
\begin{equation}
\rho=
\begin{pmatrix}
\rho_{11} & 0 & 0 & \rho_{14} \cr
0 & \rho_{22} & \rho_{23} & 0 \cr
0 & \rho_{32} & \rho_{33} & 0 \cr
\rho_{41} & 0 & 0 & \rho_{44}
\end{pmatrix}
\ ,
\label{2.1}
\end{equation}
with $\rho_{32}=\rho_{23}^*$ and $\rho_{41}=\rho_{14}^*$.
Normalization requires $\rho_{11}+\rho_{22}+\rho_{33}+\rho_{44}=1$, while positivity of $\rho$
imposes:
\begin{equation}
\rho_{ii}\geq0\ ,\ i=1,2,3,4\ , \quad
\rho_{11}\rho_{44}-|\rho_{14}|^{2}\geq0\ ,\quad
\rho_{22}\rho_{33}-|\rho_{23}|^{2}\geq0\ .
\label{2.1b}
\end{equation}
Equivalently, a two-qubit density matrix can also be represented in the so-called {\it Fano form},
a generalization of the one-qubit decomposition in (\ref{1.20}):
\begin{equation}
\rho=\frac{1}{4}\Big[ \sigma_0\otimes\sigma_0 + \sum_{i=1}^3 R_{0i}\, \sigma_0\otimes\sigma_i
+ \sum_{i=1}^3 R_{i0}\, \sigma_i\otimes\sigma_0 +\sum_{i,j=1}^3 R_{ij}\, \sigma_i\otimes\sigma_j\Big]
\ .
\label{2.2}
\end{equation}
However, only elements from the set
\begin{equation}
{\cal X}=\Big\{ \sigma_0\otimes\sigma_0,\, \sigma_0\otimes\sigma_3,\,
\sigma_3\otimes\sigma_0,\, \sigma_1\otimes\sigma_2,\, \sigma_2\otimes\sigma_1,\,
\sigma_1\otimes\sigma_1,\, \sigma_2\otimes\sigma_2,\, \sigma_3\otimes\sigma_3\Big\}\ ,
\label{2.3}
\end{equation}
should have a nonvanishing contribution in order to reproduce the X-shape in (\ref{2.1}), and one finds:
\eject
\begin{equation}
\begin{array}{ll}
R_{03}=\rho_{11} - \rho_{22} + \rho_{33} - \rho_{44}\ ,\phantom{\Big)} 
& R_{11}=\rho_{14} + \rho_{41} + \rho_{32} + \rho_{23}\ ,\\
R_{30}=\rho_{11} + \rho_{22} - \rho_{33} - \rho_{44}\ ,\phantom{\Big)}
& R_{22}=\rho_{32} + \rho_{23} - \rho_{14} - \rho_{41}\ ,\\
R_{12}=i\big(\rho_{14} - \rho_{41} + \rho_{32} - \rho_{23}\big)\ ,\phantom{\Big)} 
& R_{33}=\rho_{11} + \rho_{44} - \rho_{22} - \rho_{33}\ ,\\
R_{21}=i\big(\rho_{14} - \rho_{41} - \rho_{32} + \rho_{23}\big)\ .\phantom{\Big)} & \\
\end{array}
\label{2.4}
\end{equation}
Similarly, also the generator ${\rm id}\otimes \mathbb{L}$ of the two-qubit semigroup $\Gamma_t={\rm id}\otimes\gamma_t$
should have a specific form in order for $\Gamma_t$ to preserve the same shape;
this request puts some constraints on the generator $\mathbb{L}$ of the single-qubit dynamical map $\gamma_t=\e^{t\mathbb{L}}$.
As discussed in the previous section, $\mathbb{L}$ can be in general decomposed 
into Hamiltonian and dissipative contributions:
\begin{equation}
\mathbb{L}[\rho]= -i\big[H,\, \rho\big]
+\sum_{i,j=1}^{3}{\cal C}_{ij}\Bigl(\sigma_j\rho\,\sigma_i-
{1\over 2}\Bigl\{\sigma_i\sigma_j,\, \rho\Bigr\}\Bigr)\ ,
\label{2.5}
\end{equation}
for a generic hermitian 2-dimensional matrix $H$ and $3\times 3$ hermitian coefficient matrix ${\cal C}$. However, only when
$H$ is proportional to $\sigma_3$ and ${\cal C}$ takes the form (\ref{1.16}), 
the two-qubit density matrix will remain of the form (\ref{2.1}) under the dynamical map ${\rm id}\otimes \e^{t\mathbb{L}}$,
thus justifying the choices made in the previous section.%
\footnote{To be precise, the form (\ref{1.16}) of the Kossakowski matrix can be the result of a more general
environment than the one considered in Section 2; specifically, one can allow generic two-point correlations
for the $B_1$ and $B_2$ bath operators (with also $G_{12}\neq 0$), leaving only $B_3$ as an independent variable
($G_{13}=G_{23}=\,0$).
However, this slight generalization would not add new physical insights to the discussed results, 
while making the treatment more involved.}
This result is the direct consequence of the specific decomposition of the Lie algebra $su(4)$, as generated
by the 16 elements $\sigma_\mu\otimes\sigma_\nu$, $\mu,\nu=0,1,2,3$, induced by its subalgebra ${\cal X}$;
indeed, $su(4) = {\cal X} \oplus {\cal X}^\perp$, where ${\cal X}^\perp$ is the complement set of ${\cal X}$,
together obeying the following algebraic relations under multiplication~\cite{Quesada}:
\begin{equation}
{\cal X}\cdot{\cal X} \subset {\cal X}\ ,\quad {\cal X}^\perp\cdot{\cal X}^\perp\subset{\cal X}\ ,\quad
{\cal X}\cdot{\cal X}^\perp\subset{\cal X}^\perp\ ,\quad {\cal X}^\perp\cdot{\cal X}\subset{\cal X}^\perp\ .
\label{2.6}
\end{equation}
Because of these relations, the set ${\cal X}$ is preserved by ${\rm id}\otimes \mathbb{L}$, and consequently by
the finite dynamical map $\Gamma_t$ it generates, only for the just specified choice
of Hamiltonian and Kossakowski matrix.

An additional advantaged of the X-shape density matrix in (\ref{1.16}) is that its entanglement content
can be explicitly evaluated through the computation of its concurrence $\frak{C}[\rho]$ \cite{Bennett,Wooters}. 
Indeed, one finds:
\begin{equation}
\frak{C}[\rho] = 2\ {\rm max}\Big\{ 0,\ |\rho_{23}|-\sqrt{ \rho_{11}\rho_{44} },\ 
|\rho_{14}|-\sqrt{ \rho_{22}\rho_{33} }\Big\}\ .
\label{2.7}
\end{equation}
For sake of definiteness, in the following we shall assume to start at $t=\,0$ with an entangled state
for which $|\rho_{23}|>\sqrt{ \rho_{11}\rho_{44}}$. 

The evolution of the density matrix (\ref{2.1}) under the dynamical map $\Gamma_t={\rm id}\otimes\gamma_t$,
with $\gamma_t$ as given by (\ref{1.27}), (\ref{1.28}), can be explicitly expressed as:
\begin{eqnarray}
\nonumber
&& \rho_{11}(t) = \frac{1}{4}\bigg[\Big(1+\Lambda(t)\Big) \Big(1 + R_{30}\Big) + \e^{-2\gamma t} \Big(R_{03} + R_{33}\Big)\bigg]\ ,\\
\nonumber
&& \rho_{22}(t) = \frac{1}{4}\bigg[\Big(1-\Lambda(t)\Big) \Big(1 + R_{30}\Big) - \e^{-2\gamma t} \Big(R_{03} + R_{33}\Big)\bigg]\ ,\\
\nonumber
&& \rho_{33}(t) = \frac{1}{4}\bigg[\Big(1+\Lambda(t)\Big) \Big(1 - R_{30}\Big) + \e^{-2\gamma t} \Big(R_{03} - R_{33}\Big)\bigg]\ ,\\
\label{2.8}
&& \rho_{44}(t) = \frac{1}{4}\bigg[\Big(1-\Lambda(t)\Big) \Big(1 - R_{30}\Big) - \e^{-2\gamma t} \Big(R_{03} - R_{33}\Big)\bigg]\ ,\\
\nonumber
&& \rho_{14}(t) = \frac{1}{4}\bigg[\Big(M_{11}(t) R_{11} + M_{12}(t) R_{12} - M_{21}(t) R_{21} - M_{22}(t) R_{22}\Big)\\
\nonumber
&&\hskip 2cm -i\Big(M_{21}(t) R_{11} + M_{22}(t) R_{12} + M_{11}(t) R_{21} + M_{12}(t) R_{22}\Big)\bigg]= \rho_{41}^*(t)\ ,\\
\nonumber
&& \rho_{23}(t) = \frac{1}{4}\bigg[\Big(M_{11}(t) R_{11} + M_{12}(t) R_{12} + M_{21}(t) R_{21} + M_{22}(t) R_{22}\Big)\\
\nonumber
&&\hskip 2cm +i\Big(M_{21}(t) R_{11} + M_{22}(t) R_{12} - M_{11}(t) R_{21} - M_{12}(t) R_{22}\Big)\bigg]= \rho_{32}^*(t)\ .
\end{eqnarray}
First of all, one easily checks that the reduced density matrix for the second qubit obtained by tracing over the first one,
$\rho^{(2)}(t)={\rm Tr}_1\big[\rho(t)\big]$, remains positive for all times. Indeed, one finds:
\begin{equation}
\rho^{(2)}(t)=\frac{1}{2}\Big(\sigma_0+ R_{03}(t)\sigma_3\Big)\ ,\quad 
R_{03}(t)=-\frac{w}{\gamma}\Big( 1- \e^{-2\gamma\, t}\Big) + \e^{-2\gamma t} R_{03}\ .
\label{2.9}
\end{equation}
As $|R_{03}|\leq1$, because of the positivity of the initial state, and, recalling (\ref{1.26}),
$0\leq w/\gamma\leq 1$, one deduces that also $|R_{03}(t)|\leq1$ for all times, thus assuring 
$\rho^{(2)}(t)\geq0$, for any X-shaped initial state (\ref{2.1}).

Instead, one expects that the positivity of a generic initial state (\ref{2.1}) will not be preserved
by the evolution map ${\rm id}\otimes\gamma_t$ in (\ref{2.8}), as the Redfield dynamics $\gamma_t$
discussed in Section 2 is non-positive. In order to examine this issue in more detail, 
we shall focus on the following four-parameter family of initial two-qubit states:
\begin{equation}
\tilde\rho=
\begin{pmatrix}
\mu & 0 & 0 & u \cr
0 & \nu & iv & 0 \cr
0 & -iv & 1-2\mu-\nu & 0 \cr
u & 0 & 0 & \mu
\end{pmatrix}
\ ,
\label{2.10}
\end{equation}
where $\mu$, $\nu$, $u$ and $v$ are real constants satisfying the following inequalities, necessary for positivity:
\begin{equation}
\mu\geq0\ , \quad \nu\geq0\ ,\quad 0\leq 2\mu + \nu\leq 1\ ,\quad u^2\leq \mu^2\ ,\quad v^2\leq \nu(1-2\mu-\nu)\ .
\label{2.11}
\end{equation}
From (\ref{2.4}), one further obtains: $R_{03}=-R_{30}=1 -2(\mu+\nu)$, $R_{12}=-R_{21}=2v$, $R_{11}=-R_{22}=2u$
and $R_{33}=4\mu-1$.
As mentioned before, we shall also assume a non-vanishing concurrence, $\frak{C}[\tilde\rho]=2(|v|-\mu)>0$.

By using the explicit time evolution given in (\ref{2.8}), and taking for simplicity
the bath temperature to be zero, so that, recalling  (\ref{1.26}), $w=\gamma$, one verifies by inspection that the diagonal
elements of $\tilde\rho(t)$ remain non-negative for all times:
\begin{equation}
\begin{array}{ll}
\tilde\rho_{11}(t) = \mu\, \e^{-2\gamma t}\ ,\phantom{\Big)} 
& \tilde\rho_{33}(t) = (1-2\mu-\nu)\, \e^{-2\gamma t}\ ,\\
\tilde\rho_{22}(t) = \mu \big(1- \e^{-2\gamma t}\big) + \nu\ , \phantom{\Big)}
& \tilde\rho_{44}(t) = 1-\mu-\nu - \e^{-2\gamma t}\big(1-2\mu-\nu\big)\ .\\
\end{array}
\label{2.12}
\end{equation}
Concerning the additional quadratic inequalities in (\ref{2.1b}), recall that the bath dissipative parameters 
$a$, $b$, $\alpha$ and $\gamma$ in (\ref{1.23}) are all proportional to $\lambda^2$, which is assumed to be small. 
Consequently, one can neglect them with respect to the qubit frequency $\omega$, and,
as a result, take in practice $2\tilde\omega/\Omega\simeq 1$. Within this approximation, from (\ref{2.8}) one gets:
\begin{eqnarray}
\nonumber
&&\tilde\rho_{11}(t)\, \tilde\rho_{44}(t) -|\tilde\rho_{14}(t)|^2 = 
\mu \e^{-2\gamma t} \Big[1-\mu-\nu - \e^{-2\gamma t}\big(1-2\mu-\nu\big)\Big] - u^2\, \e^{-2(a+\alpha) t}\\
&& \hskip 4cm\geq \mu^2 \e^{-2\gamma t} - u^2\, \e^{-2(a+\alpha) t}\geq0\ ,
\label{2.13}
\end{eqnarray}
since $\mu^2\geq u^2$, with the additional assumption $a +\alpha-\gamma\geq0$. Similarly, one also finds:
\begin{eqnarray}
\nonumber
&&\tilde\rho_{22}(t)\, \tilde\rho_{33}(t) -|\tilde\rho_{23}(t)|^2 = 
(1-2\mu-\nu) \e^{-2\gamma t} \Big[\nu +\mu\big(1- \e^{-2\gamma t}\big)\Big] - v^2\, \e^{-2(a+\alpha) t}\\
&& \hskip 4cm\geq \nu(1-2\mu-\nu) \e^{-2\gamma t} - v^2\, \e^{-2(a+\alpha) t}\geq0\ ,
\label{2.14}
\end{eqnarray}
since $\nu(1-2\mu-\nu)\geq v^2$. As a consequence, the four-parameter family of density matrices in (\ref{2.10})
constitute a set of admissible initial states for the evolution ${\rm id}\otimes\gamma_t$, with $\gamma_t$ the non-positive
Redfield dynamics (\ref{1.27}), since they remain positive for all times.

Let now focus on the entanglement content of the evolving density matrix $\tilde\rho(t)$. We first consider the
behaviour of the concurrence $\frak{C}[\tilde\rho(t)]$ for small times. Using the explicit
expressions in (\ref{2.8}) expanded to first order in $t$, one finds:
\begin{equation}
\big|\tilde\rho_{23}(t)\big| - \sqrt{\tilde\rho_{11}(t) \tilde\rho_{44}(t)} =
|v|-\mu +\Big[ \gamma (3\mu+\nu-1) -\big( (a+\alpha)v +2bu\big)\Big]\ t + O(t^2)\ .
\label{2.15}
\end{equation}
It is sufficient to choose an initial state $\tilde\rho$ for which
$1-2\mu-\nu\leq\mu$, and $u \leq -(a+\alpha)v/2b$ to immediately conclude that $\frak{C}[\tilde\rho(t)]$
does increase in time as soon as the dynamics sets in, clearly an unphysical result 
for a Markovian dynamics.

A different situation occurs for non-Markovian single-qubit dyamics $\gamma_t$: in this case,
by adding a second ancillary qubit, the resulting two-qubit time evolution, again of the 
factorized form ${\rm id}\otimes\gamma_t$, might be able to increase the
entanglement between the two qubits \cite{Lofranco,Aolita}. 
As the ``true'', unapproximated reduced qubit dynamics, obtained by just tracing over the bath degrees of freedom,
is in general non-Markovian, one might then be tempted to conclude that the just signaled
increase of entanglement of the Redfield dynamics actually reproduces a real phenomenon.
In fact, this is not the case, as the initial state of the three-party qubit+ancilla+bath total system,
$\tilde\rho\otimes\rho_B$, is of the so-called ``Markov type'', and for such states 
no non-Markovian reduced two-qubit dynamics exists able to augment at any later time 
the entanglement of the initial two-qubit state \cite{Sargolzahi}.

\begin{figure}[h!]
\centering
\includegraphics[width=1\linewidth]{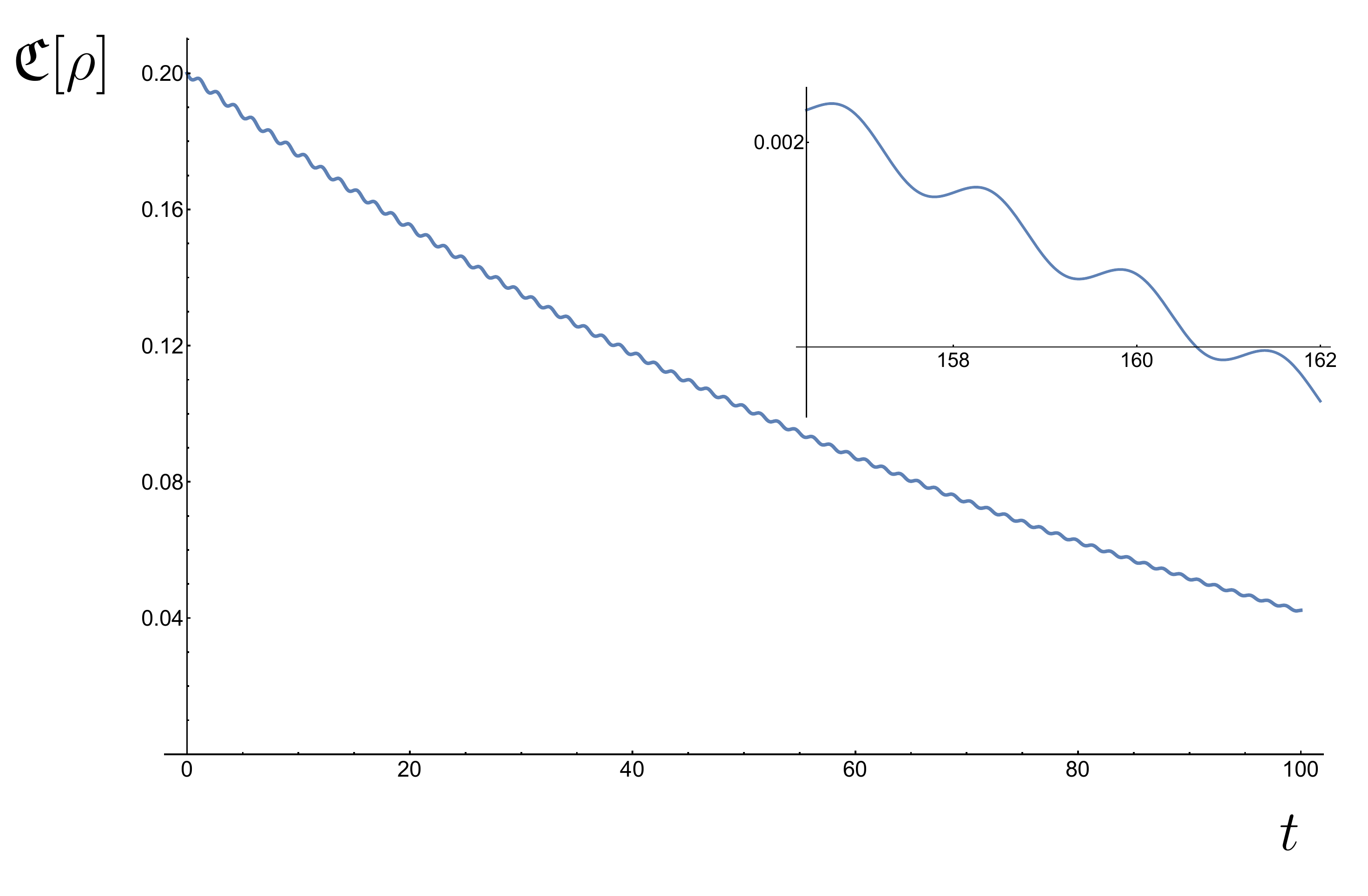}
\caption{\small\sl Behaviour of concurrence as a function of time in units $1/\omega$, with bath parameters $a/\omega=0.005$, 
$b/\omega=0.05$, $\alpha/\omega=0.001$, $\gamma/\omega=0.001$, $w/\gamma=0.5$, for an initial state with
$\mu=0.025$, $\nu=0.1$, $u=0.02$ and $v=0.125$; these values satisfy all the constraints (\ref{2.11}) assuring the positivity
of the initial density matrix, and the initial assumption of a weak coupled environment.
The insert is an enlargement of the region where concurrence becomes zero: 
the oscillatory behaviour of concurrence is not a transient phenomenon, it persists for over 160 cycles.}
\label{Fig1}
\end{figure}
%


The unphysical increase of entanglement signaled by (\ref{2.15}) is a consequence
of the non-positivity of the Redfield dynamics (see also the discussion in the next section) and as such
is not limited to small times; indeed, Fig.1 displays the behaviour of the concurrence
for an initial state belonging to the family in (\ref{2.10}), 
in a range of constants $\mu$, $\nu$, $u$ and $v$, different from that just examined in the small time regime,
whose positivity is nevertheless preserved by the ${\rm id}\otimes\gamma_t$ dynamics.
The bath is now at finite temperature, with bath parameters $a$, $b$, $\alpha$, $\gamma$, $w$ 
taken to be small with respect to $\omega$, as $\lambda\ll\omega$ by assumption; time
is measured in units of $1/\omega$. The plot shows an oscillatory shape, 
leading to a sudden death of entanglement, as expected for a damping dynamics. However, at each cycle, concurrence
increases, a clear contradiction as the dynamical map is in factorized form. This periodic increase of entanglement
is not a transient effect, disappearing after a few oscillations: it persists for very long times, as long as
concurrence remains nonvanishing.


\vskip 2cm

\section{Discussion}

In modelling the reduced dynamics of quantum systems weakly coupled to heat-baths, semigroups
of Bloch-Redfield type are often viewed as convenient choices. They are generated by master equations
that can be readily obtained through second order approximations in the system-bath coupling
and a naive Markovian limit.

However, for such dynamics, positivity of the system states is not guaranteed,
as unphysical negative eigenvalues in the reduced density matrix begin to develop as soon as
the dynamics starts. This ``shortcoming'' has not prevented the wide use 
of Redfield evolutions in applications, and various {\it ad hoc} prescriptions have been adopted
in order to deal with the appearance of ``negative probabilities''.

As the lack of positivity of the reduced density matrix occurs in general for small times,
to avoid inconsistencies, one often used prescription is to restrict the possible system initial states 
to those for which the Redfield dynamics $\gamma_t$ remains positive. Interestingly,
this prescription works also when the system under study is statistically (but not dynamically!)
coupled to another inert ancilla, provided the initial state is in separable form,
\hbox{$\rho=\sum_i p_i\, \rho_i^{(1)} \otimes \rho_i^{(2)}$}, $p_i\geq0$, $\sum_i p_i=1$,
with $\rho_i^{(1)}$, $\rho_i^{(2)}$,
admissible states for ancilla and system, respectively; indeed, in this case the time evolution
is governed by the map ${\rm id}\otimes\gamma_t$ and no negative eigenvalues can possibly develop.

Nevertheless, this prescription fails for entangled initial states, even when their positivity is
preserved by the action of ${\rm id}\otimes\gamma_t$, as further physical inconsistencies may arise.
Indeed, as shown in the previous sections, in the case of a two-qubit system, the first inert, while the second evolving
with a Redfield dynamics $\gamma_t$, their mutual entanglement can periodically increase under
the action of ${\rm id}\otimes\gamma_t$, a purely local operation. This unexpected phenomenon
is not confined to the beginning of the dynamics: on the contrary, it persists for finite times,
as long as the entanglement is not vanishing.
The physical inconsistency of Redfield dynamics is therefore not limited to
the initial occurrence of ``negative probabilities''; rather, it manifest itself at finite times
in the generation of entanglement through the action of local maps.

A similar phenomenon can be observed for the mutual information $I(S\!\!:\!\!A)$ of anciliary ($A$) + system ($S$) two-qubit model. 
This quantity  provides the information about the total correlations present in the bipartite $A+S$ system
\begin{equation}\label{}
  I(S\!:\!A)  = S(\rho_{SA} || \rho_S \otimes \rho_A) ,
\end{equation}
where $S(\rho||\sigma) = {\rm Tr}(\rho[\log\rho - \log\sigma])$ denotes a relative entropy, 
$\rho_{SA}$ is an ancilla+system  state, while
$\rho_A = {\rm Tr}_S [\rho_{SA}]$ and $\rho_S = {\rm Tr}_A [\rho_{SA}]$, 
the ancella and system reduced density matrix. For a completely positive and trace perserving (CPTP) semigroup one obviously has
\begin{equation}\label{S<0}
  \frac{d}{dt} S\big(\rho_{SA}(t) || \rho_S(t) \otimes \rho_A(t) \big) \leq 0 ,
\end{equation}
where $\rho_{SA}(t) = (e^{t \mathcal{L}} \otimes {\rm id})[\rho_{SA}]$, {\it i.e.} the total correlations present in the initial state $\rho_{SA}$ monotonically decrease. Simple analysis shows that for the Redfield dynamics the inequality (\ref{S<0}) is again violated. Moreover, it is violated even for times when the dynamics is already completely positive.  

It is worth stressing that this unphysical behaviour also affects Redfield evolutions of more general form,
of type $\gamma_t\otimes\gamma_t$. These dynamical maps describe the reduced dynamics of two equal, independent,
non-interacting systems, both immersed in a common environment; as 
$\gamma_t\otimes\gamma_t= \big({\rm id}\otimes\gamma_t\big)\circ\big(\gamma_t\otimes{\rm id}\big)$,
local generation of entanglement would also occur in this more general setting.%
\footnote{In this regard, it might be worth recalling that the dynamical map $\gamma_t\otimes\gamma_t$
is positive-preserving if and only if $\gamma_t$ is completely positive \cite{Benatti-5}.}
In addition, through straightforward extensions of the bipartite setting, 
Redfield dynamics will clearly show similar inconsistencies also
in the case of multipartite systems.

These considerations seems to suggest the presence of an intrinsic incompatibility of Redfield type dynamics 
with entangled states. As mentioned before, the origin of such inconsistency has to be found with the way
the free evolution generated by the system Hamiltonian interferes with the approximations used to derive
the Redfield dynamics. This is clearly indicated by the oscillating behaviour of concurrence in Fig.1,
with a period $1/\omega$, the inverse of the free system energy unit. The same periodicity can be found
in the time behaviour of the Choi matrix, $\big({\rm id}\otimes\gamma_t\big)[P_+]$, where $P_+=|\psi_+\rangle\langle \psi_+|$,
with $|\psi_+\rangle=(|0\rangle\otimes|0\rangle +|1\rangle\otimes|1\rangle)\sqrt{2}$, $\sigma_3|i\rangle=(-1)^{i+1}|i\rangle$,
$i=0,1$, whose positivity signals the complete positivity of the dynamical map $\gamma_t$.
For the Redfield evolution discussed above, the Choi matrix starts having negative eigenvalues as soon as
the dynamics starts, becoming positive only at longer times, after having oscillating between 
being negative and positive.

It should be stressed, however, that a Redfield semigroup $\gamma_t = e^{t\mathcal{L}}$ even when becomes completely positive is still not CP-divisible, {\it i.e.} the intermediate map (a propagator) $\gamma_{t,s} = e^{(t-s)\mathcal{L}}$ for $t> s$ need not be completely positive; actually, for small $\tau$ the map $e^{\tau\mathcal{L}}$ violates even positivity. Hence,  although 
the Redfield semigroup becomes CPTP after some finite time $T$, it still violates CP-divisibility and displays typical non-Markovian dynamical effects such as non-motonicity of concurrence and mutual information. In a sense the dynamical map for $t > T$ remembers that initially the very condition of positivity was violated.

One way to deal with all these problems is through the precise mathematical treatment devised by
Davies,%
\footnote{For a detailed derivation and discussion, {\it e.g.} see \cite{Alicki}-\cite{Rivas}.}
that in the particular case discussed in the previous sections amounts to substitute the dissipative
generator $\mathbb{D}$ in (\ref{1.6}), (\ref{1.7}), with the following ergodic average:
\begin{equation}
\widetilde{\mathbb{D}}
=\lim_{T\to+\infty}\frac{1}{2T}\int_{-T}^{+T}{\rm d}\tau\,
{\rm e}^{-\tau\mathbb{L}_S}\circ \mathbb{D}\circ
{\rm e}^{\tau\mathbb{L}_S}\ ,
\end{equation}
where $\mathbb{L}_S[\,\cdot\,]=-i\big[H_S, \cdot\,\big]$ is the generator of the free Hamiltonian
system dynamics. This average operation will then transform the $4\times 4$ matrix generator
$\mathcal{L}$ in (\ref{1.22}) into a new one $\widetilde{\mathcal{L}}$, of the same form
as (\ref{1.22}), with $b=0$ and $a$ and $\alpha$ replaced by $(a+\alpha)/2$.
As $a$, $\alpha$, $\gamma$ are positive, the family of transformations $\tilde\gamma_t$
generated by $\widetilde{\mathcal{L}}$ form a semigroup of completely positive maps,
having as asymptotic state the same Gibbs equilibrium state (\ref{1.30}) of the original Redfield evolution.
In this case, no inconsistencies will possibly arise as such dynamics are perfectly compatible
with the presence of entanglement, in all possible physical situations, thus suggesting a safer alternative
to the use of Redfield dynamics, at least within the domain of applicability of the Davies prescription \cite{Merkli}.

\end{document}